# Detection of the laser induced damage using a He-Ne laser reflective imaging technique

F. Novák[1,2,*], L. Uvarova[1,3], Š. Němcová[1,4] and M.-G. Mureșan[1]

[1] HiLASE Centrum, Institute of Physics of the Czech Academy of Sciences, Dolní Břežany, Czechia
[2] Faculty of Nuclear Sciences and Physical Engineering, Czech Technical University, Břehová 7, 115 19 Prague, Czechia
[3] Department of Surface and Plasma Science, Faculty of Mathematics and Physics, Charles University, V Holešovičkách 2, Prague 180 00, Czechia
[4] Faculty of Mechanical Engineering, Czech Technical University in Prague, Technická 4, 166 07 Prague, Czechia

*Author to whom any correspondence should be addressed.

E-mail: frantisek.novak@hilase.cz

**Abstract**

Laser-induced damage is a serious challenge for optical components; therefore, determining the laser-induced damage threshold (LIDT) is a crucial step in the manufacturing process. In many cases, such as for space applications, it is also necessary to account for the vacuum environment. Since conventional damage detection methods face limitations under vacuum conditions, an alternative approach is required. This article introduces a He–Ne laser imaging system designed for *in-situ* damage detection within a LIDT station. The system enables fourfold magnification imaging of the test sample and its surroundings without placing imaging optics inside the vacuum chamber, thereby preserving the cleanliness of the chamber. The detection method can be applied to both transparent and opaque samples; in transparent optics, damage is observable from either side. To verify the system's functionality, two sample types were investigated: a silica wafer (non-transparent for He–Ne radiation) and a commonly used dielectric mirror for 1030 nm (transparent for He–Ne radiation). Particular attention was devoted to determining the minimum damage size that can be reliably recognized. The system successfully distinguished damage features as small as 35 μm.

## 1. Introduction

The study of laser-induced damage threshold (LIDT) of optical materials plays an important role in areas such as research-grade laser optics (coating and thin-film research [1,2]), space/aerospace laser optics (study and development of their properties according to space-related conditions [3,4,5,6,7]), and medicine (skin [8,] or cornea laser thresholds [9]).

LIDT testing of optical components is also becoming critical for high-energy high-power (HEHP) laser systems development, especially for improvement of anti-reflection/high-reflection (AR/HR) coatings of laser crystals [10] and their pumping [11].

One of the critical parts of the LIDT experiments is the *in-situ* damage detection. Based on the large number of articles regarding LIDT tests, several methods of *in-situ* damage detection are employed:
- long working distance microscopy [12,13];
- fluorescence detection [14];
- scattered light monitoring [15,16];
- plasma spark monitoring [17];
- online fast camera microscopy [18];

In many cases (i.e. for space application) it is necessary to take the vacuum environment into account. Because the LIDT of materials can be dependent on the surrounding pressure, such studies are important. On the other hand, applying some conventional *in-situ* damage detection methods under vacuum conditions may present difficulties. Therefore, for LIDT measurements in vacuum, an efficient method of *in-situ* damage detection is required. There are several limitations, such as the dimensions of the vacuum chamber (either small or too large) or the difficulty of sample removal during the experiment. Thus, applying a He-Ne laser in reflection imaging mode for the *in-situ* damage detection part was proposed.

The method of damage detection based on reflected light was described in several articles [15, 19-28]. Usually, a He-Ne laser was the preferred light source, but other wavelengths could also be employed: 530 nm laser in reflected mode [15]. Liu et al. used a He-Ne gas CW laser at AOI of 30° (angle of incidence) to detect the damage to the sample [19]. At the sample the He-Ne laser beam had a diameter of 1 mm, with the reflected beam directed to a power meter. Damage occurrence was indicated by a reduction in the reflected beam power. Yang et al. [20] employed a He-Ne laser collinearly combined with the Nd:YAG testing beam for damage detection. The method relied on the scattering of the He-Ne laser light from the damage. Also, *in-situ* He-Ne detection was used in several other articles [21-27], where damage detection was based on He-Ne laser scattering or diffusion. Via scattering the smallest detection limit was reached by Kafka [27] under 1 µm.

In [29], Mikami utilized He-Ne laser imaging in a transmission setup, where the camera captured the fraction of the He-Ne radiation transmitted through the sample to study the temperature dependence of the LIDT of single-layer coatings on silica glass substrates [28, 29]. Mikami reported the ability to recognize damage with dimensions as small as 2 µm, but it is not clear from the text, whether this limit really belongs to the detection *in-situ* or *ex-situ* with Nomarski differential interference microscopy.

It is also worth mentioning the work of Bartels et al. [30], who investigated the effect of laser conditioning on the LIDT of AR optics using a chopped He-Ne laser beam for in-situ damage detection. In their setup, the He-Ne laser beam overlapped with the main UV laser beam used during the LIDT tests, and the scattered He-Ne light from damage sites was detected by a Si photodiode. The authors concluded that they were able to detect damage down to 5 µm in size.

According to the above-mentioned literature, both reflection- and transmission-based methods have been tested. It was concluded that reflected He-Ne light imaging detection is more attractive for application in our case due to the simplicity and versatility of this method.

In this work the setup for damage detection based on the reflection and imaging of the He-Ne laser radiation is presented. The process and methodology for establishing the smallest in-situ detectable damage using a He-Ne laser system are described in detail. The detection method can be applied to both transparent and non-transparent samples, and in the case of transparent optics, damage can be observed regardless of the side on which it occurs.

## 2. Experimental

This section is dedicated to the description of the He-Ne detection system and samples that were selected for these experiments – silica wafer and highly reflective (HR) dielectric plane mirror for 1030 nm. Damage at samples was either already created during the earlier experiments or during the measurent by q-switched Nd:YAG laser. All experiments were carried out at standard pressure and a temperature of 22 °C. During the measurements, the relative humidity ranged between 40 % and 55 %.

### 2.1 Experimental setup

The He-Ne laser damage detection system is shown in Figure 1 (Zemax scheme in the Supplement Figure S1). He-Ne laser radiation irradiated the sample. Part of the radiation is reflected, passes through the lens L3 and comes to the CMOS detector. The principle of the system is based on the intensity contrast development between the damaged and undamaged area of the sample. In the the damaged area large amount of the photons is scattered, resulting in the reduction of the laser intensity captured at the CMOS detector. Imaging and magnification of the sample picture is processed by the lens L3 and relative positions of the sample, lens L3 and CMOS camera.

The sample was mounted in a metal holder, which was attached to two crossed translation stages enabling precise movement of the samples during the LIDT test. The damages were induced by Nd:YAG laser (Q-smart 450, Lumibird) operating at 1064 nm. The AOI of the Nd:YAG laser was set to 3° to prevent back-reflection into the laser cavity. The laser delivered pulses with energies of up to 0.5 J and a duration of 8.5 ns. A vacuum chamber will be integrated into the LIDT station to enable testing under vacuum conditions, which is essential for evaluating space optics.

The He-Ne probe beam was aligned to overlap the Nd:YAG laser spot on the sample at an incidence angle of 15° relative to the Nd:YAG beam axis. Combined with the 3° AOI of the Nd:YAG laser, this configuration resulted in the He-Ne radiation interacting with the sample at an effective AOI of 18°. The final value of the chosen angle was primarily determined by the construction of the vacuum chamber and the overall configuration of the individual elements of the LIDT station. The general objective was to achieve the smallest possible value in order to avoid potential undesirable optical phenomena on the vacuum chamber windows (such as astigmatism).

The core component of the imaging system was a linear S-polarized 632.8 nm He-Ne laser (HNLO50L, Thorlabs) with a beam diameter of 0.81 mm (1/e²). The beam diameter at the sample plane could be changed by adjusting the distance between the two plano-convex lenses L1 (f = 60 mm) and L2 (f = 75 mm). Damage imaging was performed using a bi-convex lens (L3) and a CMOS camera (Beamage-3.0, Gentec) with a 2048 × 1088 pixels sensor of physical dimensions 11.3 × 6.0 mm² and a pixel pitch of 5.5 × 5.5 μm². The required distances between lens L3 and the sample (denoted $a$) and between lens L3 and the camera (denoted $a'$) were calculated using the following equations to achieve the desired magnification:

$$\frac{1}{a} + \frac{1}{a'} = \frac{1}{f} \quad (1)$$

$$Z = -\frac{a'}{a} \quad (2)$$

where $f$ denotes the focal length of the lens, $a$ is the object distance, $a'$ is the image distance and $Z$ is the magnification. In addition, a 632.8 nm bandpass filter was mounted in front of the CMOS camera to reduce the noise. The He-Ne laser was chosen as the probe beam source because its wavelength differs from those of the Nd:YAG fundamental, second, and third harmonics. This spectral separation ensured that only the probe beam reached the sensor, effectively filtering out any residual radiation from the pump laser.

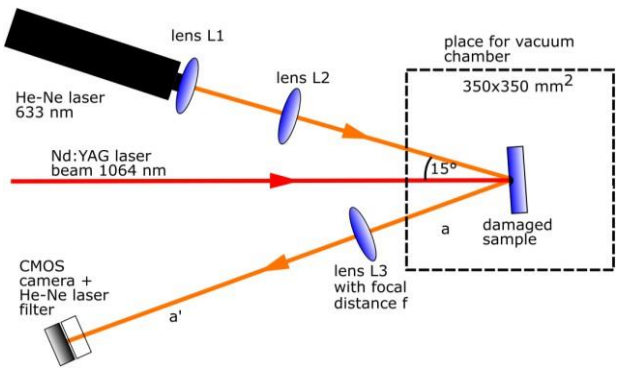

**Figure 1.** Experimental LIDT setup, using He-Ne laser damage detection. L1, L2 – illuminating lenses, L3 – imaging lens with focal distance f, a – object distance, a´ - image distance.

There were several limiting factors in our setup. The internal dimensions of the vacuum chamber will be equal to 350 × 350 × 350 mm³. The sample will be located at its center. In order to minimize the number of optical elements inside the vacuum chamber, lens L3 should have been placed outside the chamber. Consequently, the distance between the sample and the imaging lens had to be at least (350/2)×(1/cos(15°))=182 mm.

The second limiting factor was the overall compactness of the setup. The distance between the sample and the camera did not have to exceed the distance between the He-Ne laser and the sample, which was approximately 1300 mm. To achieve maximum magnification while respecting this spatial constraint, lens L3 with a focal length of f = 200 mm was selected. As a result the magnification of |Z| = 4 was reached, which means according to the equation (1) a distance of $a$=250 mm and consequently $a'$=1000 mm. Lens L3 was mounted on a translation rail to achieve fine adjustment along the optical axis between the sample and the camera.

Based on the magnification |Z| = 4, pixel size 5.5 × 5.5 μm² and assuming that effective noise suppression requires the dark spot area on the camera sensor to span at least five pixels in one dimension, a detection limit of about 7 μm for the smallest observable damage was established.

Another limiting factor of the detection system arises from Rayleigh's criterion for resolving two-point sources based on their diffraction patterns. The minimum resolvable feature size r in a diffraction-limited system is given by

$$r = 1.22 \frac{\lambda \cdot a}{D} \cong 4 \, \mu m, \quad (3)$$

where λ is the wavelength of the probe beam - λ=632.8 nm, $a$ is the distance between the sample and the imaging lens - $a$=250 mm, and $D$ is the aperture diameter of the lens – D=50.8 mm. A potential improvement in resolution would be to use a lens with a shorter focal length, allowing the lens according to the Equation (1) to be placed closer to the sample and thereby reducing the value of $a$ at constant $D$. However, this solution is restricted by the physical dimensions of the vacuum chamber, which limits the minimum achievable distance between the sample and the detection limit of the CMOS camera.

## 2.2 Samples

To verify the functionality of the detection setup and determine the smallest detectable damage site, two samples were used: a silicon wafer with dimensions of 43 × 43 mm² and a thickness of 0.525 mm, one side polished and a dielectric mirror with one side coated (HR@1030 nm, AOI 0°, BK7 substrate). The diameter of the mirror was 1" (25.4 mm) and the thickness 9 mm.

The selection of these two samples was motivated by their differing optical properties. The refractive index of silicon at a wavelength of 632.8 nm is 3.88 [31]. At an incidence angle of 18°, this results in a reflectivity of approximately 37% for

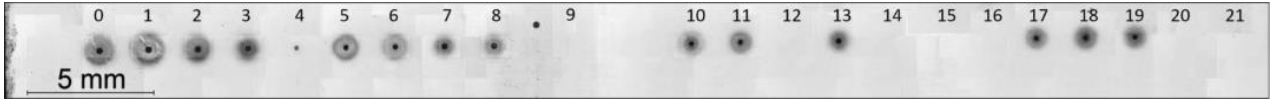

**Figure 2.** Silicon wafer sample map of the newly created damage sites (captured by confocal microscopy). Some of the damages are hardly observable because of the absence of typical colored area surrounding the damage.

S-polarized light at the air–silicon interface; the remaining light is absorbed or scattered.

The dielectric mirror was highly reflective at its design central wavelength of 1030 nm. However, for He-Ne laser radiation, the coated (front) surface was only partially reflective. In contrast to the silicon wafer, the absorption of He-Ne radiation in the BK7 substrate is minimal. As a result, part of the light is reflected from the rear surface at the BK7–air interface. This back-reflected radiation passes again through the front surface and interferes with the portion of the beam reflected from the front coating. Such interference gives rise to fringes in the recorded images, which may affect the visibility of laser-induced damage.

Both samples had previously been damaged by laser exposure during earlier experiments. The laser-induced damage spots on both samples were spaced at regular intervals. These equidistant positions were used to aid in the identification of the corresponding damage sites in the camera images. Nevertheless, both samples still contained sufficiently large undamaged areas suitable for inducing new, controlled laser damage.

New laser-induced damage sites were also identified during detection based on their relative positions, determined from the LIDT test parameters – specifically, the fixed spacing between individual sites. Additionally, the identification process was aided by reference damage marks presented on each sample. Measurements of the damage morphology were conducted using a confocal microscope (OLS5000, Olympus).

## 3. Results of experiments and their discussion

This section presents the experimental results of *in-situ* detection using a He-Ne laser on a silicon wafer and a dielectric mirror, both in real-time and non-real-time modes. For the silicon wafer, only non-real-time *in-situ* detection was carried out, while for the dielectric mirror both real-time and non-real-time detection were applied. In the non-real-time case, two different configurations were used for the dielectric mirror, depending on the side of the damage occurrence and the side from which the He-Ne beam was incident. The detectability of the damage was evaluated by a human observer, while computer-based image analysis of the CMOS camera data will be addressed in future work.

### 3.1 Damage detection on the silicon wafer sample

On the silicon wafer sample, 22 new laser damage sites of various sizes were created and analyzed (Figure 2), along with 6 pre-existing damage sites from earlier experiments, as summarized in Supplement table S1 (important spots in Table 1). The new damages were created using the Nd:YAG laser focused by a lens with a focal length of f = 500 mm. The beam exhibited a near-Gaussian spatial profile with a diameter of 169 μm ($1/e^2$) at the sample surface, an ellipticity of 98%. Fluence levels were selected based on prior LIDT testing of similar samples. The resulting damage size depended on both the laser fluence and the number of pulses delivered. The 6 additional damage sites originating from earlier experiments were also created using the same Nd:YAG laser.

**Table 1.** List of important spots studied during the experiments with He-Ne laser detection and their parameters by the silica wafer sample. Spots 10,21 was newly damaged, while spots 25,26 were previously damaged. The dimensions of damages were denoted by Δx and Δy.

| Spot # | Fluence (J/cm²) | Pulses | Δx (μm) | Δy (μm) | He-Ne Detection |
|---|---|---|---|---|---|
| 10 | 3.6 | 100 | 165 | 155 | yes |
| 21 | 3.0 | 250 | 60 | 55 | yes |
| 25 | – | – | 35 | 30 | yes |
| 26 | – | – | 5 | 10 | no |

The smallest damage created using a fluence of 3.0 J/cm² had dimensions of 60 × 55 μm² (spot 21). From the previously tested sites, damage of the dimensions 35 × 30 μm² (spot 25) and one smaller damage (5 × 10 μm², spot 26) were included in this work. As shown in Table 1 and Supplement Table S1, the He-Ne detection setup successfully detected damage at all locations except for the smallest site (spot 26).

Figure 3a presents a typical damage site (spot 10) induced by the Nd:YAG laser, with dimensions of 165 × 155 μm². Its corresponding He-Ne image in Figure 3b reveals morphological features surrounding the damaged region. Figure 3c presents the smallest damage produced using 3.0 J/cm² fluence. The corresponding He-Ne detection is shown in Figure 3d. The figures 3e and 3f show the damage in spot 25 – the smallest detectable damage with the dimensions 35x30 $\mu m^2$. The tiny damage on spot 26 can be seen in Figure 3g. Only a weak diffraction pattern was visible at the location of spot 26 (see Figure 3h) while using the He-Ne detection, resulting in a detection limit which can be estimated at around 35 $\mu m$ on silicon sample.

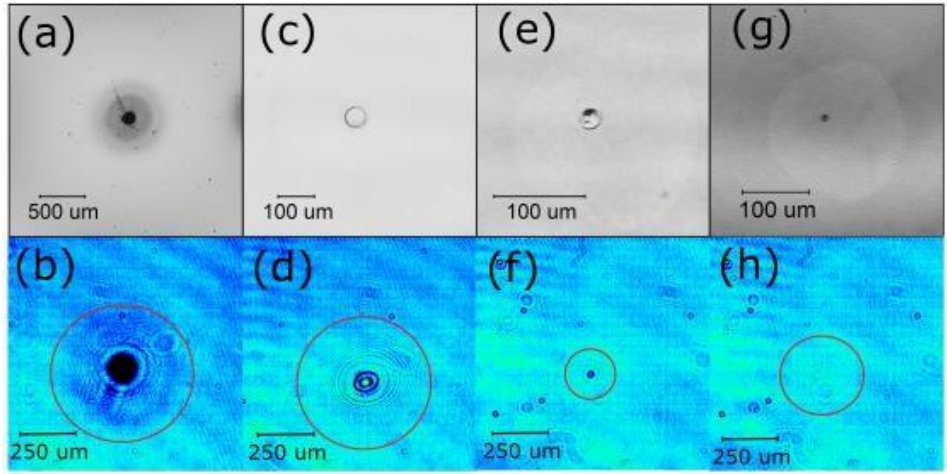

**Figure 3.** Damage at spots 10 (a and b), 21 (c and d), 25 (e and f), and 26 (g and h) of the silicon wafer sample as captured by confocal microscopy and CMOS camera under He-Ne laser illumination, respectively.

**Figure 4.** Map of the damaged spots in the (a) coated and (b) uncoated side of the dielectric mirror newly created by Nd:YAG laser (captured by confocal microscopy).

For smaller damage sizes, there is a risk of misidentifying the damage as a dust particle or another form of contamination. In the case of in-situ real-time measurements, such confusion can be excluded by comparing images taken before and after laser irradiation. However, for in-situ non-real-time measurements, this comparison is not possible. Nevertheless, the known positions of the damage sites can be used for identification, since the sites were arranged at equidistant intervals. Furthermore, the origin of a given image feature can be verified by slightly translating the sample: the image of a dust particle on the optical components or detector will remain stationary, whereas the image of an actual damage site will shift together with the sample. This method, however, does not apply to dust particles located directly on the sample surface.

### 3.2 Damage detection on the dielectric mirror

As with the silicon wafer sample, both newly induced and previously existing damage sites were analyzed on the dielectric mirror. A total of 12 new damage sites were created by a Nd:YAG laser (see Figure 4). The beam exhibited a near-Gaussian spatial profile with a diameter of 165 μm (1/e²) at the sample, an ellipticity of 97%. The laser beam was focused on the coated side of the mirror (S1). In four cases, damage was also observed on the uncoated side (S2). With the exception of spot 3 (see Supplement table Su2), all S2 side damage occurred at higher laser fluence levels, which can be explained by the high laser-induced damage threshold (LIDT) of the substrate material. With regard to spot 3, the statistical nature of the damage threshold must be considered.

The smallest recorded damage occurred at a fluence of 76.7 J/cm² with 10 pulses in spot 4 (on S1), resulting in dimensions of 50 × 35 μm². With the exception of spot 5, all S2 side damage sites exhibited larger lateral dimensions than those on the S1 side. In addition, two complex damage sites from previous experiments were analyzed to verify the detection capabilities of the system (see Figures 6e and 6g). These damages were created using the BIVOJ laser system (flat top spatial profile, 1030 nm, 2-14 ns, energy up to 7 J). Details regarding the BIVOJ laser system and the LIDT testing procedure used were previously published [32,33].

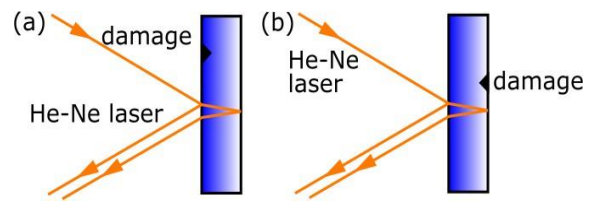

**Figure 5.** Damage detection setups used for the transparent optics - (a) damage at the front side (DFS) and (b) damage at the rear side (DRS).

As with the silicon wafer, damage detection using the He-Ne probe laser was performed *in-situ*, but not in real-time mode. However, unlike the case of the silicon wafer, detection was performed in two distinct configurations, realized by rotating the sample by 180 degrees within the holder.:

- Damage at the front side (DFS): the He-Ne beam was incident from the same side where the damage occurred.
- Damage at the rear side (DRS): the He-Ne beam was incident from the opposite side than the damage occurred.

This two-configuration approach was motivated by the fact that during LIDT testing laser-induced damage may also occur on the rear surface of the sample — opposite to the laser entry side (Figure 5). All results of the He-Ne laser-based detection and corresponding damage parameters are summarized in Supplement Table S2 (important spots in Table 2).

For the DFS configuration, the damages captured by confocal microscope and their images at CMOS camera are depicted in the Figure 5. Typical damage site created at a fluence of 2.8 J/cm² is shown in Figure 6a. The damage shape closely followed the Gaussian profile of the laser beam, with an approximate diameter of 135 μm.

The corresponding He-Ne laser image is shown in Figure 6b, where the previously discussed interference pattern (arising from reflections at both the S1 and S2 sides) is clearly visible. Based on image analysis, the average spacing between interference maxima is approximately 550 μm, while the average width of the destructive interference band (i.e. the dark fringe) is about 170 μm. Considering that the imaging system provides a magnification of $|Z|=4$, this implies that detection of damage smaller than approximately $170\ \mu m / 4 \approx 40\ \mu m$ may be challenging. The smallest laser-induced damage observed during the experiments is shown in Figure 6c and its corresponding He-Ne image in Figure 6d (spot 4). Rear-side damage is visible in the right portion of Figure 6d, where it reveals as a localized disturbance or compression of the interference field. Figures 6f and 6h present complex damage structures from earlier experiments. In both cases, the damaged regions are clearly recognizable. For spots 12F and 13C, however, it remains uncertain whether such damage would remain visible if it was located within a region of destructive interference. On the other hand, even small defects can slightly disrupt the destructive interference pattern, as demonstrated in the cases of damage sites 12D, 13A, and 13B (see Figures 6f and 6h), which may be considered a benefit of using coherent probe radiation. Thus, the smallest reliably detectable damage size under the given imaging conditions is approximately 45 μm.

**Table 2** List of important spots studied during the experiments with He-Ne laser detection and their parameters by the dielectric mirror. Spots 0-11 were newly damaged, while spots 12-13 were previously damaged. The detection column indicates whether the damage is detected in the DFS setup or in the DRS setup. The dimensions of damages were denoted by $\Delta x$ and $\Delta y$. Damage side S1 – coated, S2 - uncoated.

| Spot # | Fluence (J/cm²) | Pulses | $\Delta x$ (μm) | $\Delta y$ (μm) | Damaged Side | He-Ne Detection | |
|---|---|---|---|---|---|---|---|
| | | | | | | DFS setup | DRS setup |
| 0 | 49.3 | 10 | 160 | 145 | S1 | yes | no |
| 4 | 76.7 | 10 | 50/445 | 35/605 | S1+S2 | yes + yes | no + yes |
| 7 | 42.8 | 100 | 105 | 115 | S1 | yes | yes |
| 8 | 42.8 | 50 | 130 | 135 | S1 | yes | yes |
| 11 | 58.1 | 20 | 150 | 165 | S1 | yes | no |
| 12A | – | – | 70 | 40 | S1 | yes | no |
| 12B | – | – | 55 | 50 | S1 | yes | no |
| 12C | – | – | 250 | 320 | S1 | yes | yes |
| 12D | – | – | 80 | 100 | S1 | yes | yes |
| 12E | – | – | 45 | 50 | S1 | yes | no |
| 12F | – | – | 35 | 40 | S1 | yes | no |
| 13A | – | – | 130 | 85 | S1 | yes | yes |
| 13B | – | – | 50 | 30 | S1 | yes | no |
| 13C | – | – | 30 | 25 | S1 | yes | yes |
| 13D | – | – | 95 | 40 | S1 | yes | no |

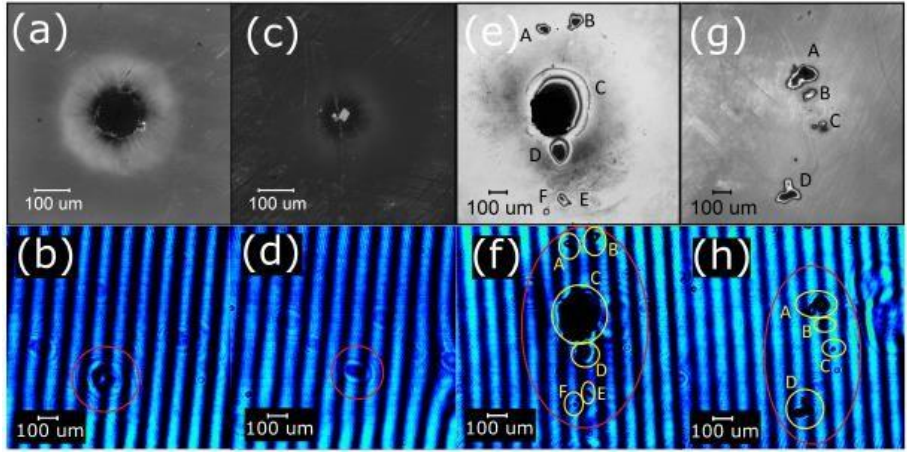

**Figure 6.** Damages at spots 8 (a and b), 4 (c and d), 12 (e and f), and 13 (g and h) of the dielectric mirror as captured by confocal microscopy and CMOS camera under He-Ne laser illumination, respectively.

Detection of laser-induced damage in the DRS configuration is generally more challenging than DFS configuration. In the DFS configuration, the damage appears as a dark region within the interference field, as it scatters light reflected from both sides. In contrast, in the DRS configuration, light reflected from the front side is unaffected by damage, because its location is in the rear side. The damage instead reveals as a local disturbance of the interference pattern — typically seen as a weakening of interference maxima, an enhancement of minima, and the emergence of localized fringe structures. Nonetheless, when compared to the regular interference field of an undamaged region, such disruptions can still be well distinguished, which may be considered a benefit of using coherent probe radiation (for instance, coherent light is the basis of an electronic speckle pattern interferometry (ESPI) technique, which is used for material defect detection [34]). An additional advantage of employing coherent radiation is the possibility of identifying whether the damage is situated on the front or back side based on the image of the damage. It is also important to consider that if damage occurs at corresponding positions on both the front and rear surfaces, the rear-side damage image will appear laterally shifted relative to the front-side damage image due to geometric optical effects. Therefore, the angle of incidence of the laser beam must be selected carefully to avoid spatial overlap between rear-side and front-side damages—particularly when these are produced at regular intervals. In the performed experiment, this spatial displacement led to the non-detection of damage at spots 0 and 11 - in both cases, the shifted image of the rear-side damage would have appeared in a region obscured by the sample mechanical holder. The damage site at spot 8 (marked by a red circle) is shown in Figure 7a as a representative example of damage at S1 side imaged using the damage at rear side (DRS) detection setup. The damage is visible as a disruption of the interference field. In contrast, the detection of the 50 × 35 μm damage at spot 4 (also S1-side damage) was debatable in the DRS configuration. Figures 7c and 7d present complex damage sites. In the case of spot 12, multiple damage sites (a-e)—can be clearly identified. For spot 13, only damage sites 13A and 13D are recognizable. Among all identified sites, spot 12E represents the smallest reliably detectable damage, with dimensions of 45 × 50 μm. The smallest damage site reliably detected in this configuration was spot 7, with dimensions of 105 × 115 μm (see Figure 8b). Therefore, the smallest reliably detectable damage size in the DRS configuration can be estimated at approximately 50 μm, which is comparable to the detection threshold observed in the DFS configuration.

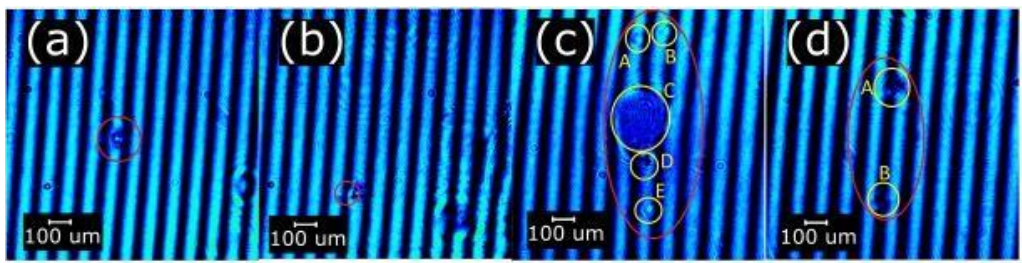

**Figure 7.** Damages at spots 8 (a), 4 (b), 12 (c) and 13 (d) of the dielectric mirror illuminated with He-Ne laser on CMOS camera in DRS configuration.

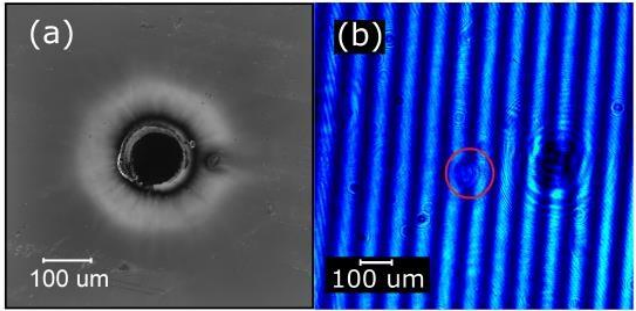

**Figure 8.** Damage at spot 7 of the dielectric mirror as captured by confocal microscopy (a) and CMOS camera under He-Ne laser illumination, respectively (b).

### 3.3 In-situ damage detection in real-time

*In-situ* in real-time damage detection was performed only on a dielectric mirror during LIDT testing, as above described. As previously noted, *in-situ* detection in real-time offers the advantage of allowing direct comparison between images recorded before and after the laser exposure. This contrast significantly improves damage identification. Figures 9a and 9b show spot 7 before and after the laser shot, respectively. In the figure 9c there is a clear observed damage, based on the subtraction of the figures 9a and 9b. A limitation of *in-situ* in real-time detection, however, is the inability in some cases to predict which surface of the sample (front or rear) will be damaged. Because only one surface can typically be imaged in focus, damage on the opposite side may not be clearly resolved. This effect has to be taken into account during real-time detection but it can be overcome by sufficiently high depth of field of the imaging optics.

### 3.4 Discussion

The He–Ne-based detection setup demonstrated relatively strong detection capabilities; however, several limitations must be taken into account. First, a fundamental restriction arises for samples that do not sufficiently reflect light at the He–Ne laser wavelength, for instance materials that appear 'black' to the human eye and are used in the production of astronomical instruments [35]. For such materials, different acceptable wavelengths can be applied within the same detection system configuration. Second, detection in the case of thinner dielectric mirrors may be more difficult. Since the distance between interference fringes is inversely proportional to the optical thickness of the substrate, thinner mirrors are expected to produce more widely spaced interference patterns. This results in a reduced spatial resolution of interference-related contrast and may worsen the detection threshold for small damage sites. On the other hand, since damage disrupts the interference structure, this effect may not pose a significant problem. Furthermore, the interference structure could be tuned, although only to a limited degree, by adjusting the angle of incidence of the He–Ne laser radiation.

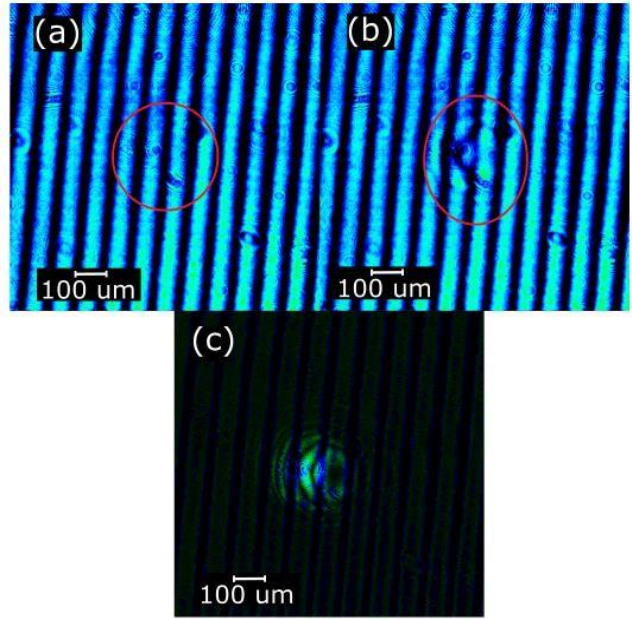

**Figure 9.** He-Ne laser detection image of spot 7 of the dielectric mirror before (a) and after (b) damage creation. At (c) subtraction of the images.

The most similar detection setup is described in [29]. The author utilized He–Ne laser imaging in a transmission setup, where the camera captured the fraction of the He–Ne radiation transmitted through the sample. The author reported the ability to recognize damage with dimensions as small as 2 μm, but, as it already has been mentioned, it is not clear, whether the detection limit really belongs to the detection *in-situ* or *ex-situ* with Nomarski differential interference microscopy. Based on the facts that the magnification was equal to 3 and CCD camera was used, more likely this was not the limit of in-situ detection using the He–Ne laser. In addition, unlike this transmission configuration, our reflection configuration offers the advantage that detection can be performed even on opaque samples.

Overall, in-situ detection methods based on scattered He–Ne light appears capable of detection of smaller damage sizes (e.g., <1 μm [27] and 5 μm [30]). However, these methods provide only limited information on the morphology and dimensions of the damage, carries the risk of misinterpreting contamination particles as damage, and does not allow effective imaging of the broader sample area.

With our method, the detection limit for the silicon wafer was established at 35 μm. A 5 × 10 μm defect could not be detected, most likely due to limitations of the CMOS camera (pixel size) combined with interference effects from contaminants on the sample or optics. The shallow depth and weak contrast of the damage site may also have contributed. As mentioned earlier, higher magnification could be achieved if the imaging lens were positioned inside the chamber; alternatively, the same lens could provide higher magnification by increasing the distances $a$ and $a'$. However this would imply operating with a larger beam incident on the imaging lens andlarger $a'$ conflicting with the compactness requirements of the setup. Another important limitation arises from the resolution of the CMOS camera and the diffraction limit given by the Rayleigh criterion. Nevertheless, future improvements are expected to push these limits and enable detection of even smaller damage features.

## 4. Summary

In this work the setup for damage detection based on the reflection and imaging of the He-Ne laser radiation was presented. To assess the system's performance, two types of samples were tested: a silicon wafer and a dielectric mirror. For the silicon wafer, the minimum detectable damage size was 35 μm. In the case of the dielectric mirror, two detection configurations were tested: damage at the front side (DFS) and damage at the rear side (DRS), which differed by the position of the damage relative to the incident He-Ne laser radiation. The smallest reliably detectable damage size was found to be 45 μm in the DFS and 50 μm in the DRS configuration, respectively.

## Data availability statement

All data that support the findings of this study are included within the article (and any supplementary files).

## Acknowledgements

This work was co-funded by the European Union and the state budget of the Czech Republic under the project LasApp CZ.02.01.01/00/22_008/0004573.

The authors would like to thank Pavel Crha (HiLASE Engineering and Technical Support) for his contribution during the setup realization, designing and manufacturing custom mechanics.

## Conflict of interest

The authors declare that they have no known competing financial interests or personal relationships that could have appeared to influence the work reported in this paper.

## References

1. Wang, Y.; Cheng, X.; Shao, J.; Zheng, C.; Chen, A.; Zhang, L. 2022 The Damage Threshold of Multilayer Film Induced by Femtosecond and Picosecond Laser Pulses. *Coatings* **12** 251
2. Velpula, Praveen Kumar, Daniel Kramer, and Bedrich Rus. 2020 Femtosecond Laser-Induced Damage Characterization of Multilayer Dielectric Coatings *Coatings* **10** 603.
3. J. Cao, T. Chen, H. He, Y. Xu, Y. Chen, N. Gao, Y. Du, Z. Wu, Y. Liu, 2023 Visible light performance of x-ray filters for Einstein probe *Optical Engineering* **62** 044103.
4. P. Weßels, A. Büttner, M. Ernst, M. Hunnekuhl, R. Kalms, L. Willemsen, D. Kracht, J. Neumann 2018 UV-DPSS laser flight model for the MOMA instrument of the ExoMars 2020 Mission, in: Z. Sodnik, N. Karafolas, B. Cugny (Eds.) *International Conference on Space Optics — ICSO 2018*, Vol. **11180**, International Society for Optics and Photonics, SPIE, 2019 111801B
5. W. Riede, H. Schroeder, G. Bataviciute, D. Wernham, A. Tighe, F. Pettazzi, J. Alves 2011 Laser-induced contamination on space optics, in: G. J. Exarhos, V. E. Gruzdev, J. A. Menapace, D. Ristau, M. J. Soileau (Eds.), *Laser-Induced Damage in Optical Materials: 2011* **8190** 81901E.
6. D. Wernham, J. Alves, F. Pettazzi, A. P. Tighe, 2010 Laser-induced contamination mitigation on the ALADIN laser for ADM-Aeolus, in: G. J. Exarhos, V. E. Gruzdev, J. A. Menapace, D. Ristau, M. J. Soileau (Eds.), *Laser-Induced Damage in Optical Materials: 2010* **7842** 78421E
7. C. Y. Sheng, 2006 Effects of laser-induced damage on optical windows in the presence of adhesives under simulated thermal-vacuum conditions, in: G. J. Exarhos, A. H. Guenther, K. L. Lewis, D. Ristau, M. J. Soileau, C. J. Stolz (Eds.), *Laser-Induced Damage in Optical Materials: 2006* **6403** 64030H
8. Jean M, Schulmeister K. 2021 Laser-induced injury of the skin: validation of a computer model to predict thresholds. *Biomedical Optics Express*. **12** 2586-2603

9. Mathieu Jean, Karl Schulmeister, David J. Lund, and Bruce E. Stuck, 2021 Laser-induced corneal injury: validation of a computer model to predict thresholds *Biomed. Opt. Express* **12**, 336-353
10. H. Wang, A. R. Meadows, E. Jankowska, E. Randel, B. A. Reagan, J. J. Rocca, C. S. Menoni, 2020 Laser induced damage in coatings for cryogenic yb:yag active mirror amplifiers, *Opt. Lett.* **45** 4476–4479
11. J. Chen, H. Lin, D. Hao, Y. Tang, X. Yi, Y. Zhao, S. Zhou, 2019 Exaggerated grain growth caused by zro2 doping and its effect on the optical properties of tb3al5o12 ceramics, *Scripta Materialia* **162** 82–85
12. H. Wang, A. R. Meadows, E. Jankowska, E. Randel, B. A. Reagan, J. J. Rocca, C. S. Menoni, 2020 Laser induced damage in coatings for cryogenic yb:yag active mirror amplifiers, *Opt. Lett.* **45** 4476–4479
13. J. Neauport, N. Bonod, S. Hocquet, S. Palmier, G. Dupuy 2010 Mixed metal dielectric gratings for pulse compression, *Opt. Express* **18** 23776–23783
14. H. Schröder, W. Riede, E. Reinhold, D. Wernham, Y. Lien, H. Kheyrandish 2007 In situ observation of UV-laser-induced deposit formation by fluorescence measurement, in: G. J. Exarhos, A. H. Guenther, K. L. Lewis, D. Ristau, M. J. Soileau, C. J. Stolz (Eds.), *Laser-Induced Damage in Optical Materials: 2006*, **6403** 64031K
15. Willemsen, 2022 Large area ion beam sputtered dielectric ultrafast mirrors for petawatt laser beamlines, *Opt. Express* **30** 6129–6141
16. S. Fourmaux, J. Kieffer 2021 Laser induced damage threshold and incubation effects of high-power laser system optics *Quantum Electronics* **51** 751
17. O. A. Nassef, H. E. Elsayed-Ali 2005 Spark discharge assisted laser induced breakdown spectroscopy *Spectrochimica Acta Part B: Atomic Spectroscopy* **60** 1564–1572
18. G. Liu, D. Kuang 2024 Damage mechanism in plasma evolution of nanosecond laser-induced damage of germanium sheets in air and vacuum *Optics and Laser Technology* **174** 110689
19. G. Liu, D. Kuang, L. Song, C. Xu, C. Yan, 2023 Mechanism in damage variation of nanosecond laserinduced damage of germanium sheets in vacuum *Optics Laser Technology* **157** 108663
20. L. Yang, X. Xiang, X. Miao, Z. Li, L. Li, X. Yuan, G. Zhou, H. Lv, X. Zu 2015 Influence of oil contamination on the optical performance and laser induced damage of fused silica *Optics Laser Technology* **75** 76–82.
21. Y. Lien, E. Reinhold, D. Wernham, M. Endemann, M. Jost, E. Armandillo, W. Riede, H. Schröder, P. Allenspacher, 2006 Risk mitigation in spaceborne lasers, *Proc SPIE* **6190** 9
22. W. Riede, P. Allenspacher, D. Wernham, 2007 Laser qualification testing of space optics *Proceedings of SPIE* **01** 6403
23. P. Allenspacher, W. Riede, D. Wernham 2005 Vacuum laser damage test bench, *Proceedings of SPIE* **5991** 599128–1
24. C. Gingreau, T. Lanternier, L. Lamaignère, T. Donval, R. Courchinoux, C. Leymarie, J. Néauport 2018 Impact of mechanical stress induced in silica vacuum windows on laser-induced damage *Opt. Lett.* **43** 1706–1709
25. K. Juškevičius, R. Buzelis, G. Abromavicius, R. Samuilovas, S. Abbas, A. Belosludtsev, R. Drazdys, S. Kicas 2017 Argon plasma etching of fused silica substrates for manufacturing high laser damage resistance optical interference coatings *Optical Materials Express* **7**
26. J. Oulehla, J. Lazar 2014 Station for LIDT tests of optical components under cryogenic conditions, in: G. J. Exarhos, V. E. Gruzdev, J. A. Menapace, D. Ristau, M. Soileau, D. Ristau (Eds.), *Laser-Induced Damage in Optical Materials: 2014* **9237**, 923720
27. K. 3598 R. P. Kafka, N. Talisa, G. Tempea, D. R. Austin, C. Neacsu, E. A. Chowdhury 2016 Few-cycle pulse laser induced damage threshold determination of ultra-broadband optics *Opt. Express* **24** 28858–28868
28. K. Mikami, S. Motokoshi, M. Fujita, T. Jitsuno, K. A. Tanaka 2011 ``, in: G. J. Exarhos, V. E. Gruzdev, J. A. Menapace, D. Ristau, M. J. Soileau (Eds.), *Laser-Induced Damage in Optical Materials: 2011* **8190** 81900A
29. K. Mikami, S. Motokoshi, T. Somekawa, T. Jitsuno, M. Fujita, K. A. Tanaka 2013 A theoretical analysis for temperature dependences of laser-induced damage threshold, in: G. J. Exarhos, V. E. Gruzdev, J. A. Menapace, D. Ristau, M. Soileau (Eds.), *Laser-Induced Damage in Optical Materials: 2013* **8885** 88851T
30. N. Bartels, P. Allenspacher, W. Riede 2018 Laser conditioning of UV anti-reflective optical coatings for applications in aerospace, in: C. W. Carr, G. J. Exarhos, V. E. Gruzdev, D. Ristau, M. Soileau (Eds.), *Laser-Induced Damage in Optical Materials 2018: 50th Anniversary Conference* **10805** 108051Q
31. M. N. Polyanskiy 2025 Refractive index database, https://refractiveindex.info/, accessed: 2025-0629 (2025).
32. J. Pilar, M. De Vido, M. Divoky, P. Mason, M. Hanus, K. Ertel, P. Navratil, T. Butcher, O. Slezak, S. Banerjee, et al. 2018 Characterization of bivoj/dipole 100: Hilase 100-j/10-hz diode pumped solid state laser *Solid State Lasers XXVII: Technology and Devices* **10511** 120–126

33. M. G. Muresan, P. Cech, V. Bilek, P. Horodyska, D. Rostohar, A. Lucianetti, T. Mocek 2018 Laser induced damage in optical glasses using nanosecond pulses at 1030 nm in: C. W. Carr, G. J. Exarhos, V. E. Gruzdev, D. Ristau, M. Soileau (Eds.), *Laser-Induced Damage in Optical Materials 2018*: *50th Anniversary Conference* **10805** 108052A
34. Esteban Andrés Zarate, Edén Custodio G, Carlos G. Treviño-Palacios, Ramón Rodríguez-Vera, Hector J. Puga-Soberanes 2005 Defect detection in metals using electronic speckle pattern interferometry *Solar Energy Materials and Solar Cells* **88** 217-225
35. Marshall, Jennifer L., Patrick Williams, Jean-Philippe Rheault, Travis Prochaska, Richard D. Allen, and D. L. DePoy 2014 Characterization of the reflectivity of various black materials *Proceedings of SPIE* **9147** 91474F-91474F-8

# Detection of the laser induced damage using a He-Ne laser reflective imaging technique: Supplement 1

F. Novák[1,2,*], L. Uvarova[1,3], Š. Němcová[1,4] and M.-G. Mureşan[1]

[1] HiLASE Centrum, Institute of Physics of the Czech Academy of Sciences, Dolní Břežany, Czechia
[2] Faculty of Nuclear Sciences and Physical Engineering, Czech Technical University, Břehová 7, 115 19 Prague, Czechia
[3] Department of Surface and Plasma Science, Faculty of Mathematics and Physics, Charles University, V Holešovičkách 2, Prague 180 00, Czechia
[4] Faculty of Mechanical Engineering, Czech Technical University in Prague, Technická 4, 166 07 Prague, Czechia

*Author to whom any correspondence should be addressed.

E-mail: frantisek.novak@hilase.cz

---

This supplement provides additional tables of damage sites produced on silicon wafer and HR dielectric mirror, as well as a schematic of experimental setup generated using ZeMax software.

## He-Ne laser damage detection setup - Zemax model

He-Ne laser damage detection setup – model in Zemax software.

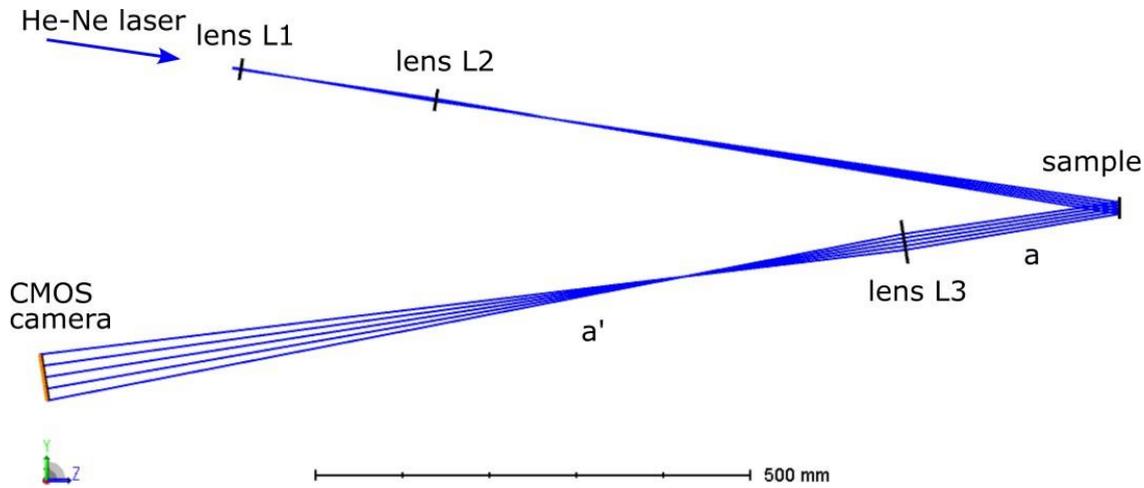

**Figure S1:** Experimental LIDT setup, using He-Ne laser damage detection – model in Zemax. L1, L2 – illuminating lenses, L3 – imaging lens, a – object distance, a´ - image distance

## Damage sites characterization of the silicon wafer sample

List of damage sites at silicon wafer sample with their dimensions characterization (by confocal microscope, Olympus OLS5000, Olympus) and detection status.

**Table S1**: Overview of fluence, number of pulses, damage dimensions, and He-Ne laser detection by the silicon wafer. Spots 0-21 were newly damaged, while spots 22-27 were previously damaged. The dimensions of damages were denoted by $\Delta x$ and $\Delta y$.

| Spot # | Fluence (J/cm$^2$) | Pulses | $\Delta x$ ($\mu$m) | $\Delta y$ ($\mu$m) | He-Ne Detection |
|---|---|---|---|---|---|
| 0 | 6.0 | 30 | 185 | 190 | yes |
| 1 | 6.0 | 20 | 185 | 185 | yes |
| 2 | 6.0 | 10 | 195 | 195 | yes |
| 3 | 6.0 | 5 | 210 | 200 | yes |
| 4 | 6.0 | 1 | 125 | 125 | yes |
| 5 | 5.0 | 30 | 190 | 185 | yes |
| 6 | 5.0 | 20 | 170 | 165 | yes |
| 7 | 5.0 | 10 | 180 | 160 | yes |
| 8 | 5.0 | 5 | 165 | 170 | yes |
| 9 | 5.0 | 1 | 70 | 80 | yes |
| 10 | 3.6 | 100 | 165 | 155 | yes |
| 11 | 3.6 | 75 | 140 | 150 | yes |
| 12 | 3.6 | 50 | 100 | 105 | yes |
| 13 | 3.6 | 25 | 155 | 160 | yes |
| 14 | 3.6 | 10 | 95 | 105 | yes |
| 15 | 3.6 | 5 | 80 | 80 | yes |
| 16 | 3.0 | 200 | 120 | 110 | yes |
| 17 | 3.0 | 500 | 170 | 155 | yes |
| 18 | 3.0 | 400 | 150 | 150 | yes |

| 19 | 3.0 | 300 | 150 | 155 | yes |
| 20 | 3.0 | 200 | 95 | 85 | yes |
| 21 | 3.0 | 250 | 60 | 55 | yes |
| 22 | – | – | 75 | 65 | yes |
| 23 | – | – | 95 | 85 | yes |
| 24 | – | – | 70 | 65 | yes |
| 25 | – | – | 35 | 30 | yes |
| 26 | – | – | 5 | 10 | no |
| 27 | – | – | 130 | 105 | yes |

## Damage sites characterization of the dielectric mirror

List of damage sites at HR dielectric mirror sample with their dimensions characterization (by confocal microscope, Olympus OLS5000, Olympus), side occurrence, setup configuration (DRS – damage at the rear side, DFS – damage at the front side) and detection status.

**Table S2**: Overview of fluence, number of pulses, damage dimensions, and He-Ne laser detection by the dielectric mirror. Spots 0-11 were newly damaged, while spots 12-13 were previously damaged. The detection column indicates whether the damage is detected in the DFS setup or in the DRS setup. The dimensions of damages were denoted by $\Delta x$ and $\Delta y$. Damage side S1 – coated, S2 - uncoated.

| Spot # | Fluence (J/cm²) | Pulses | $\Delta x$ ($\mu m$) | $\Delta y$ ($\mu m$) | Damaged Side | He-Ne Detection DFS setup | DRS setup |
|---|---|---|---|---|---|---|---|
| 0 | 49.3 | 10 | 160 | 145 | S1 | yes | no |
| 1 | 69.0 | 10 | 180 | 175 | S1 | yes | yes |
| 2 | 69.0 | 10 | 170/405 | 165/355 | S1+S2 | yes + yes | yes + yes |
| 3 | 47.1 | 10 | 545 | 425 | S2 | yes | yes |
| 4 | 76.7 | 10 | 50/445 | 35/605 | S1+S2 | yes + yes | no + yes |
| 5 | 99.3 | 10 | 220/120 | 240/95 | S1+S2 | yes + yes | yes + yes |
| 6 | 44.9 | 100 | 155 | 145 | S1 | yes | yes |
| 7 | 42.8 | 100 | 105 | 115 | S1 | yes | yes |
| 8 | 42.8 | 50 | 130 | 135 | S1 | yes | yes |
| 9 | 58.1 | 100 | 175 | 155 | S1 | yes | yes |
| 10 | 58.1 | 50 | 150 | 140 | S1 | yes | yes |
| 11 | 58.1 | 20 | 150 | 165 | S1 | yes | no |
| 12A | – | – | 70 | 40 | S1 | yes | no |
| 12B | – | – | 55 | 50 | S1 | yes | no |
| 12C | – | – | 250 | 320 | S1 | yes | yes |
| 12D | – | – | 80 | 100 | S1 | yes | yes |
| 12E | – | – | 45 | 50 | S1 | yes | no |
| 12F | – | – | 35 | 40 | S1 | yes | no |
| 13A | – | – | 130 | 85 | S1 | yes | yes |
| 13B | – | – | 50 | 30 | S1 | yes | no |
| 13C | – | – | 30 | 25 | S1 | yes | no |
| 13D | – | – | 95 | 40 | S1 | yes | no |